\documentclass[aps, prl, amsmath, showpacs, preprintnumbers, superscriptaddress, twocolumn, amssymb]{revtex4-2}
\usepackage{graphicx}
\usepackage{mathptmx, textcomp}
\usepackage[utf8]{inputenc}
\usepackage{color}
\usepackage{amssymb}
\usepackage{upgreek}
\usepackage[svgnames,x11names,table]{xcolor}
\usepackage{multirow}
\usepackage[colorlinks=true,urlcolor=DarkBlue,citecolor=DarkBlue,linkcolor=DarkBlue]{hyperref}

\newcommand{\vib}{\text{\usefont{OML}{lmr}{m}{it}\symbol{118}}}

\usepackage{calrsfs}
\usepackage{ulem}
\DeclareMathAlphabet{\pazocal}{OMS}{zplm}{m}{n}

\usepackage{ragged2e}

\def\dd{\mathrm{d}}

\begin{document}
\title{Precise photoexcitation measurement of Tan's contact in the entire BCS-BEC crossover}
\author{Manuel Jäger}
\author{Johannes Hecker Denschlag}
\affiliation{Institut für Quantenmaterie, Universität
 Ulm, 89069 Ulm, Germany}

\date{\today}

\begin{abstract}
We study two-body correlations in a spin-balanced ultracold harmonically trapped Fermi gas of $^6$Li atoms in the crossover from the Bardeen-Cooper-Schrieffer (BCS) to the Bose-Einstein-Condensate (BEC) regime. For this, we precisely measure Tan's contact using a novel method based on photoexcitation of atomic pairs, which was recently proposed by Wang \textit{et al.} [Phys. Rev. A \textbf{104} 063309 (2021)]. We map out the contact in the entire phase diagram of the BCS-BEC crossover for various temperatures and interaction strengths, probing regions in phase-space that have not been investigated yet. Our measurements reach an uncertainty of $\approx 2 \%$ and thus represent a precise quantitative benchmark. We compare our data to theoretical predictions and interpolations and localize the regions in phase space where the latter give valid results. In regions where the contact is already well known we find excellent agreement with our measurements. Thus, our results demonstrate that photoinduced loss is a precise probe to measure quantum correlations in a strongly interacting Fermi gas.
\end{abstract}

\maketitle
\justifying
A strongly interacting Fermi gas of ultracold atoms is an excellent platform for studying pair correlations and superfluidity. In the crossover between the BCS and the BEC regime, the Fermi gas undergoes a transition between two very different physical systems. In the weakly interacting BCS regime, weakly-bound Cooper pairs form on the surface of the atomic Fermi sea with its strong interparticle correlations, while in the BEC regime fermionic atoms combine to form tightly-bound bosonic molecules with small correlations between them. Tan's contact, first introduced by Shina Tan in 2008 \cite{Tan1,Tan2,Tan3}, 
 is a measure for short-range two-body correlations. From a thermodynamical point of view the total contact $\pazocal{I}$ is an extensive quantity, linearly scaling with the system size. It appears in a number of important thermodynamic relations for a strongly interacting Fermi gas. 
 
The contact and several thermodynamic relations have been investigated experimentally in various approaches, including radio-frequency (RF) spectroscopy \cite{Jin1, Jin3, Mukherjee1}, mapping of the momentum distribution \cite{Jin1,Shkedrov1}, Bragg spectroscopy \cite{Hoinka1} and inelastic decay in a Bose-Fermi mixture \cite{Laurent1}.
 In recent years, contact measurements reached uncertainties as low as two percent \cite{Hoinka1,Mukherjee1,Shkedrov1}. 
 So far, however, contact investigations were only carried out in particular areas of phase space, i.e. close to unitarity and at the lowest temperatures. A precise and comprehensive study across the entire phase diagram of the BCS-BEC crossover, providing a full picture of the contact, has been missing. 
Tan's contact is expected to change smoothly from the BCS to the BEC limit, but precise calculations of the contact are still challenging especially in the regime of strong, near resonant interactions between the particles. Therefore, precise measurements in this area will result in an important step forward towards a complete understanding of the crossover physics.
 
Here, we provide a high precision measurement of Tan's contact  across the full phase diagram of the BCS-BEC crossover for temperatures up to two times the Fermi temperature $T_F$. Due to a careful calibration the data reach an uncertainty of $\approx$ 2\% and therefore represent a quantitative benchmark to test theoretical model predictions.
For the contact measurements, we demonstrate yet another method which is based on photo-induced loss in the atomic gas, and has been outlined in \cite{Hu}. The method is closely related to early measurements of the closed-channel fraction in the group of R. Hulet \cite{Hulet1} and recent experiments in the group of  J.-W. Pan \cite{JWP}. The measured closed-channel fraction in \cite{JWP} indicated a deviation by a factor of three from the well-understood theoretical predictions on the BCS side. This deviation does not occur in our measurements.

\textit{Photoinduced two-body loss.---}As pointed out in \cite{Braaten1,Braaten2} the total contact $\pazocal{I}$ is directly linked to the inelastic two-body loss rate of a Fermi gas
\begin{equation}
-\frac{\dd N}{\dd t} = \frac{- \hbar \mathrm{Im}[a] }{2 \pi m |a|^2}  \pazocal{I},
\label{eq:lossrate}
\end{equation}
where $N$ is the total atom number, $m$ is the atomic mass and $a$ is the scattering length. 
If $a$ has a finite imaginary part this indicates two-body loss. Such loss can be induced, e.g. via photoexcitation towards an electronically excited tightly-bound molecular state. According to \cite{Hu}, in the vicinity of a magnetically tunable, intrinsically lossless Feshbach resonance, the photoinduced loss rate can then be expressed as
\begin{align}
-\frac{\dd N}{\dd t} =  \frac{\hbar \, \pazocal{I} }{2 \pi m a_{bg} W}  \frac{\Omega^2 / (2 \gamma)}{\left[1 - a_{bg}/a_s\right]^{-2} + \left[\Omega^2 / (2 \gamma \, W)\right]^2 }. \label{eq:loss}
\end{align}
Here, $\gamma$ is the linewidth of the excited molecular state and $\Omega$ is the Rabi frequency of the photoexcitation transition between the bare closed-channel bound state (of the Feshbach resonance) and the excited molecular state. $a_s$ denotes the real-valued scattering length without the photoexcitation coupling and $a_{bg}$ the corresponding background scattering length. Thus, the total contact of the spin-balanced Fermi gas can be simply deduced from the induced two-body loss rate. This is quite intuitive as two-body loss goes naturally along with two atoms being at close range.

\textit{Experiment.---}For our measurements we use an ultracold Fermi gas of $^6$Li atoms in the two lowest hyperfine states $|F = 1/2, m_F = 1/2>$ and $|F = 1/2, m_F = -1/2>$ with $N/2$ atoms per spin state. The atoms are trapped in a 3D harmonic trap which consists of a combination of an optical dipole trap and a magnetic trap. 
The atom cloud is cigar shaped, corresponding to the trapping frequencies $\omega_{ax} = 2\pi \times 21$ Hz in  axial and $\omega_{r} = 2\pi \times 150-2000$ Hz in radial direction, respectively. Using forced evaporative cooling at a magnetic field of around $790$ G we set a precise atomic temperature in the range of $0.04 - 2 $ $T_F$ for clouds with atom numbers between  \mbox{$ 5 \times 10^5$ and $ 2 \times 10^6$}. For the harmonically trapped gas, the Fermi temperature is given by $T_F = E_F / k_B = \hbar\left(\omega_{ax} \omega_r^2 3 N\right)^{1/3}/k_B$, where $E_F$ is the Fermi energy and $k_B$ is the Boltzmann constant.
By tuning the magnetic field $B$, we control the particle interaction with the help of the broad s-wave Feshbach resonance at \mbox{$832.2$ G } \cite{Hutson1}, which allows for entering both the BCS ($a_s < 0$) and BEC regimes  ($a_s > 0$) of the crossover. For this resonance the width is $W = - 2 \mu_B \times 262.3\, \mathrm{G} = - 2 \pi \hbar \times 734$ MHz, and $a_{bg} = -1582\,a_0$ where $a_0$ is the Bohr radius.
 
To optically induce two-body loss, we excite atom pairs, bound or unbound, to a deeply-bound molecular level with vibrational quantum number $\vib' = 68$  in the electronically excited state $A^1 \Sigma_u^+$ with a linewidth of $\gamma = 2 \pi \times 11.8$ MHz. For this, we make use of the fact that the initial atom pair wavefunction has an admixture
 from the (bare) molecular state $X^1 \Sigma_g^+ (\vib = 38)$, from which the state $A^1 \Sigma_u^+ (\vib' = 68)$ can be reached via an electric dipole transition \cite{Hulet1}. The photoexcitation scheme is shown in Figure S1 of the Supplemental Material \cite{Supple}.
  To drive the transition we employ a 673 nm laser beam with an intensity of a few $\upmu$W/cm$^2$ (where $\Omega \lesssim 2 \pi \times 1$ MHz), which leads to a decay of the total atom number $N$ within a few hundred milliseconds. This slow decay ensures that the system stays in thermal equilibrium during the exposure. This is in contrast to previous experiments of ours where we used fast loss to measure the pair fraction in the Fermi gas, see \cite{TP1}. We use high-field absorption imaging to measure the number of the remaining atoms, bound or unbound, as described in \cite{TP1,DH2}. Pairs that had been previously photoexcited to the molecular bound state are not detected because they quickly decay to states that do not respond to our absorption imaging scheme.\\
\begin{figure}[h!]
 \includegraphics[width=0.45\textwidth] {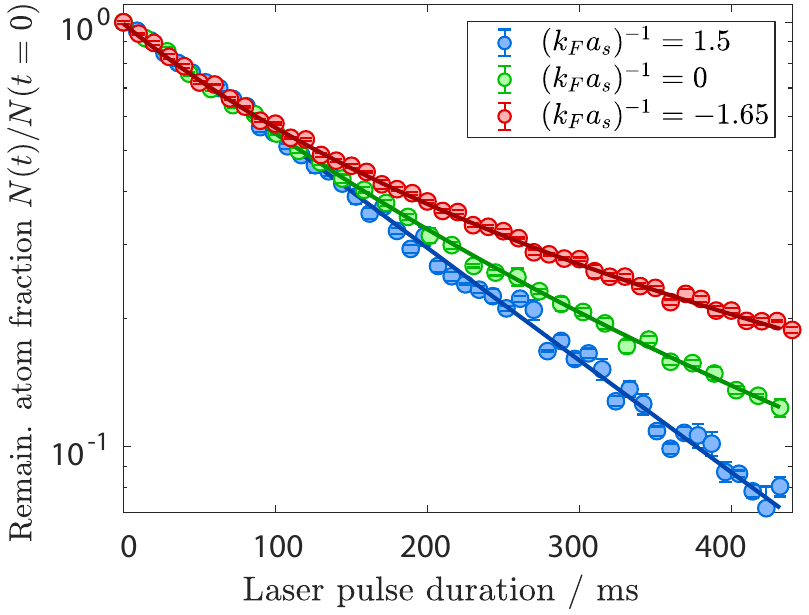}
 \vspace{-2mm}
 \caption{\justifying Remaining atom fraction as a function of the photoexcitation laser pulse duration. The measurements were carried out at magnetic fields of 753 G, 832 G and 1078 G with the initial $(k_F a_s)^{-1} = 1.5,\, 0,\,  -1.65$ corresponding to the BEC, unitarity and BCS regimes of the crossover. The continuous red and green lines are fits according to Eq. \eqref{eq:decay}, while the blue line is an exponential (i.e. $b = \infty$).}
 \label{fig1}
\end{figure}
In Fig.~\ref{fig1} we show on a logarithmic scale the remaining atoms as a function of time. The three data sets are recorded at 753 G, 832 G and 1078 G with initial interaction parameters $(k_F a_s)^{-1}$ = $1.5,\, 0$ and $-1.65$, corresponding to the BEC, unitarity and BCS regimes, respectively. Here $k_F = \left(2 m E_F / \hbar^2\right)$\textsuperscript{1/2}  is the Fermi momentum. For better comparison, the data were normalized to the initial atomic numbers $N(t = 0)$. The laser power was adjusted so that the initial relative loss rates are the same. Further details of the measurement parameters can be found in the Supplemental Material \cite{Supple}. While on the BEC side at $(k_F a_s)^{-1} = 1.5$ the loss is well described by an exponential, it is clearly non-exponential on resonance and in the BCS regime. This behavior was predicted theoretically \cite{Werner1} and 
also studied recently in \cite{JWP}. The exponential decay is typical for a pure, weakly-interacting, molecular ensemble which is present in the BEC limit at zero temperature. At unitarity and on the BCS side the non-exponential decays reflect the internal changes of the degenerate Fermi gas for different densities. 

Using Eq. \eqref{eq:loss} and the respective density-dependence of the contact in the BCS, unitarity and BEC limit one can show \cite{Werner1,Supple} that the decays at zero temperature can be described by
\begin{equation}
N(t) = N_0 / (1 + \Gamma_0 t / b)^b
\label{eq:decay}
\end{equation}
with the initial decay rate $\Gamma_0$ and $b = 2$ in the BCS limit, $b = 6$ at unitarity and $b \rightarrow \infty$ in the BEC limit.
Our fits to the data in Fig.~\ref{fig1} yield $b = 1.6 \pm 0.2$ at $(k_F a_s)^{-1} = -1.65$ and $b = 3.9 \pm 0.9$ at $(k_F a_s)^{-1} = 0$. For the decay curve at $(k_F a_s)^{-1} = 1.5$ we find that a pure exponential (or any $b \gtrsim 20$) fits well.
Despite the quantitative deviations, this shows that we already have a qualitative understanding of the decay. The deviations might be explained by a slight increase of the atom gas temperature for long photoexcitation laser pulse durations (for further discussion see \cite{Supple}).
However, we note that it is $\Gamma_0$, rather than $b$, which is relevant for the determination of the contact $\pazocal{I}$, since the  decay rate at $t = 0$, according to  Eq. \eqref{eq:decay}, is simply $\dot{N}=-N_0 \Gamma_0$. At $t = 0$, the atom number is the highest and therefore uncertainties are the smallest. This has advantages compared to other methods that rely on measuring the tails of RF-spectra or momentum distributions, where  atomic signals are generally low. In order to get precise results we accurately determine the atom numbers, the trapping frequencies, the magnetic fields and the corresponding scattering lengths, as explained in the Supplemental Material \cite{Supple}. In the following we investigate the contact in the entire BCS-BEC crossover, first for $T \approx 0$ and afterwards also for $T$ up  to 2 $T_F.$

\textit{Contact in the zero temperature limit.---}
For  $T \approx 0$ there exist already some experimental data and calculations of the contact from other groups which we can use for comparison with our results. In our measurements we typically achieve temperatures of $T < 0.04 \, T_F \lesssim T_C$ where $T_C$ is the critical temperature for superfluidity. According to Eq.~\eqref{eq:loss} we need to measure $\dd N/ \dd t$ and $\Omega$ in order to determine the contact $\pazocal{I}$. We extract the initial decay rate $\dd N/ \dd t$ from  decay curves which are similar to those shown in Fig.~\ref{fig1}. $\Omega$ is given by $\Omega^2 = k \, I$, where $I$ is the photoexcitation laser intensity and $k$ is a constant which is independent of the magnetic field $B$ and therefore of $a_s$. Following \cite{JWP}, we can conveniently determine $k$ by measuring $\dd N / \dd t$ for a given $I$ at an interaction regime where the contact $\pazocal{I}$ is known. Concretely, we chose $(k_F a_s)^{-1} \gtrsim 1$ where the contact approaches the analytical result $\pazocal{I}/N k_F = 4 \pi / k_F a_s$ \cite{Werner1}. With this calibration we can then determine $\pazocal{I}$ from measurements of $\dd N / \dd t$ at any $(k_F a_s)^{-1}$ throughout the crossover.

\begin{figure}[h!]
 \hspace{-3mm} \includegraphics[width=0.48\textwidth] {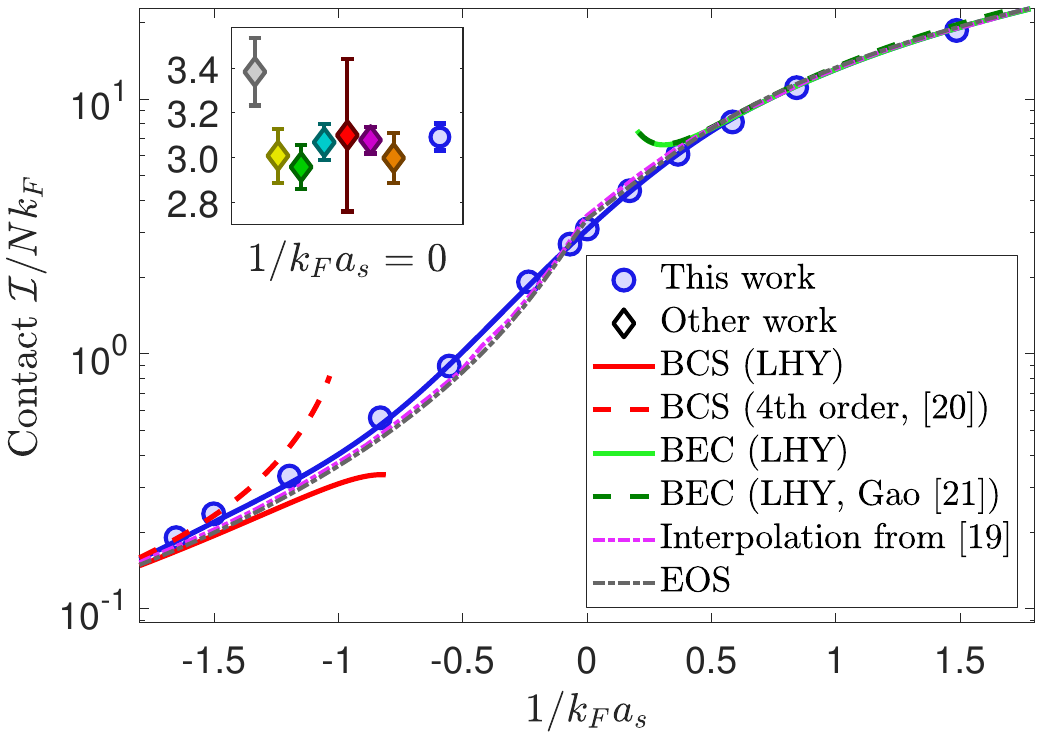}
 \vspace{-2mm}
 \caption{ \justifying Normalized contact $\pazocal{I} / N k_F$ of a harmonically trapped Fermi gas in the crossover from the BCS to the BEC regime at $T \approx 0$.
Our data (blue circles) are shown together with a guide to the eye (blue line). Uncertainties are smaller or comparable to the size of the markers.  
Also shown are trap-integrated calculations of the contact based on different approaches (see text) as well as an interpolation \cite{Werner1} (purple). The inset shows our data point (blue circle) at unitarity  and data from other groups (diamonds), namely the EOS measurement \cite{Navon1} (grey), Bragg spectroscopy measurements by \cite{Kuhnle1} (yellow) and \cite{Hoinka1} (cyan), a Quantum Monte Carlo calculation \cite{Drut1} (green), an inelastic decay measurement \cite{Laurent1} (red), and RF spectroscopy measurements \cite{Mukherjee1} (purple) and a momentum distribution measurement \cite{Shkedrov1} (orange).}
 \label{fig2}
\end{figure}

Our results are shown in Fig.~\ref{fig2} along with theoretical calculations based on ground state energy expansions in the BCS and BEC regimes (see \cite{Werner1,Supple}).
The solid red and dashed red lines are based on expansions up to second order (fermionic Lee-Huang-Yang correction) and up to the fourth order \cite{Baker1}, respectively. The results apparently converge for $(k_F a_s)^{-1} < -1.5$. The solid green and dashed dark green lines are based on the expansions to second order (bosonic Lee-Huang-Yang correction) in the BEC regime. While for the green solid line the binding energy of a dimer is calculated via $E_B = -\hbar^2 / m a_s^2$, a more accurate binding energy formula is employed for the dashed dark green line \cite{Gao1}. This leads to a $2\%$ ($4\%$) larger total contact at $(k_F a_s)^{-1} = 1 \, (2)$.

In the inset we show a comparison to other measurements and calculations at unitarity where we also find excellent agreement.
In order to compare results for homogeneous systems at unitarity with values for the contact for harmonically trapped ensembles, we divided the homogeneous results by the factor
 $(C/n k_F^\mathrm{hom})/(\pazocal{I}/N k_F) = (105 \pi / 256)  \xi^{1/4} = 1.003$ \cite{Werner1},
using $\xi = 0.367$ for the Bertsch Parameter \cite{Jensen1,Li1}. Here, $C$ is the homogeneous contact density, $n$ is the atom density and $k_F^\mathrm{hom}$ is the homogeneous Fermi momentum (see also \cite{Supple}).
  We further compare our data to calculations based on the equation of state (EOS) measurements \cite{Navon1} (see \cite{Supple}) and an interpolation from \cite{Werner1}. Here, we find small deviations in the region $-1 < (k_F a_s)^{-1} < - 0.2$ where we obtain slighly higher values for the contact.

The contact is closely related to the closed-channel fraction in the 
scattering state of two particles. In the Supplementary Material \cite{Supple} we discuss this relation and compare various experimental and theoretical studies of the closed-channel fraction. The results partially differ substantially from each other.

\textit{Finite temperature contact.---} We now perform measurements at various temperatures and couplings to map out the contact in the entire phase diagram of the BCS-BEC crossover. For this, we vary the temperature of our atom cloud between 0.04 and 2 $T_F$ by changing the  depth of our dipole potential for forced evaporative cooling. As a result we end up with around $5 \times 10^4$ ($2 \times 10^6$) atoms at our coldest (hottest) temperatures. To tune the interaction we set the magnetic field to values between 703 G and 1080 G leading to couplings in the range $-1.5 < (k_F a_s)^{-1} < 2.5$.\\
\begin{figure}[h!]
 \includegraphics[width=0.48\textwidth] {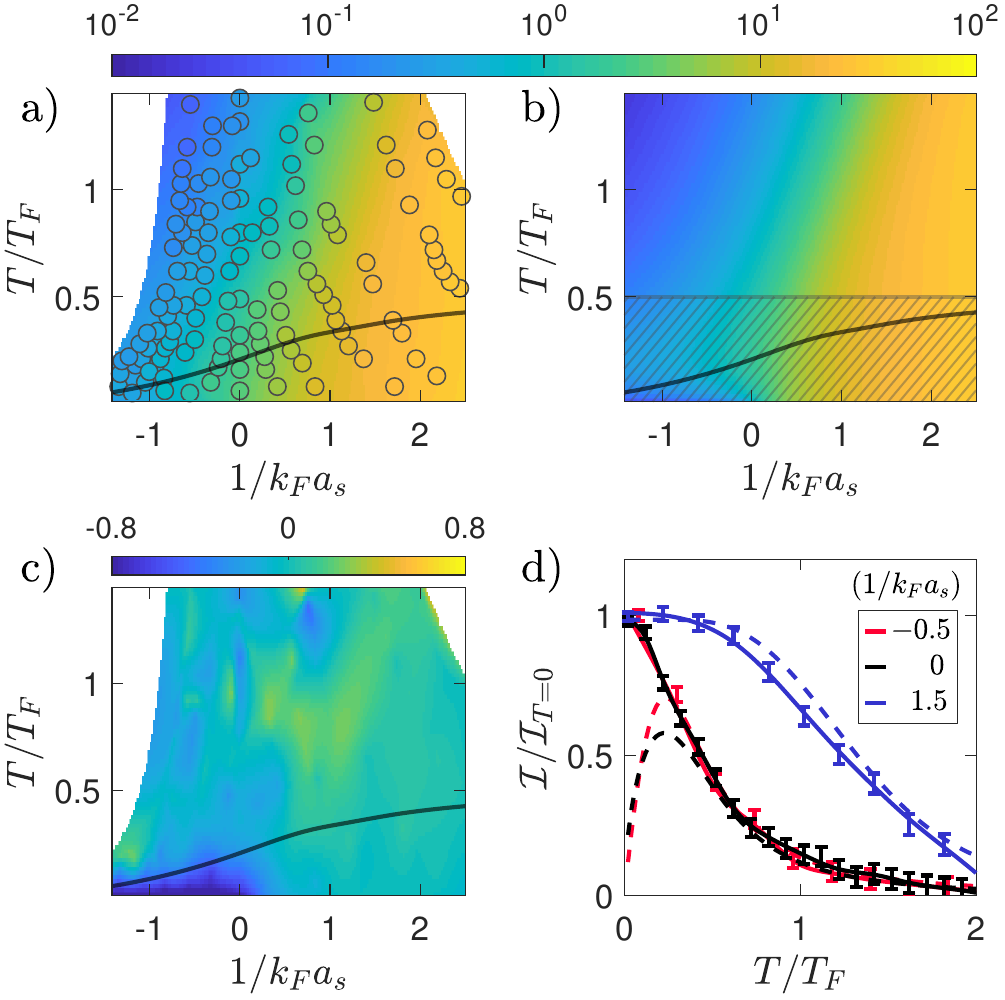}
 \vspace{-5mm}
 \caption{\justifying Map of the contact in the BCS-BEC crossover. a) Colored circles are measurements for the contact $\pazocal{I} / N k_F$, where the values are indicated by the color bar. The colored background area is an inter- and extrapolation of the measured data. The continuous black line marks the critical temperature for superfluidity $T_C$, taken from \cite{DH1}. b) The contact $\pazocal{I}_{QV,2} / N k_F$ calculated from the second-order quantum virial expansion. The shaded area below $T = 0.5 \, T_F $ marks the region where the virial expansion is expected to lose its validity. c) Relative difference  $(\pazocal{I}_{QV,2} - \pazocal{I})/ \pazocal{I}$ of our measurements and the second order quantum virial calculation. d) Contact $\pazocal{I}$ normalized by the corresponding measured zero-temperature contact as a function of temperature for three different couplings $(k_F a_s)^{-1}$. The data points are interpolations from the measured data  in a). The continuous lines are guides to the eye. The dashed lines are quantum virial calculations taken from b).}
 \label{fig3}
\end{figure}
Our measurement results are shown as colored circles in Fig.~\ref{fig3}a). Since the contact changes by three orders of magnitude within the investigated range of temperatures and couplings we plot the results logarithmically. By interpolating the data, we obtain a continuous map of the contact. Close inspection shows that this map consists of slanted, parallel stripes of color. This indicates that the description of the map might be simplified  within the given range. Indeed, as shown in the Supplemental Material \cite{Supple}, to a first approximation one can effectively replace the 2D map  by a 1D function. Although this observation is interesting, at this point we cannot offer a simple physical explanation for this.

To compare our measurements to theoretical predictions we calculated the contact within the quantum virial expansion \cite{XLiu} as done in \cite{Hu1} at unitarity. These calculations are shown in Fig.~\ref{fig3}b) and described in detail in the Supplemental Material \cite{Supple}.
As the quantum virial expansion is a series expansion in the fugacity $z =\exp(\mu / k_B T )$, it is valid at high temperatures and low chemical potentials $\mu$. The calculations show that for a harmonically trapped system the virial expansion should give valid results for the contact for temperatures as low as $T = 0.5\, T_F$,  since in this regime the fugacity is small. Below this temperature the contact values calculated with the second and third order expansion start to deviate from each other, as already discussed in \cite{Hu1}.

Fig.~\ref{fig3}c) is the relative difference between experimental data and the second order virial calculation. It shows that our measurements are in good agreement with the calculations in the given range of validity at temperatures above 0.5 $T_F$ throughout the entire crossover. On the BEC side our results agree well even down to the lowest measured temperatures of 0.04 $T_F$. This can be expected since the major contribution to the contact arises from the binding energy of the dimers, which is included in the second order virial expansion. At unitarity and on the BCS side for low temperatures, the Fermi gas is a many-body system. Since the second-order virial expansion effectively only considers interactions between two bodies, it fails to describe these regimes quantitatively. Furthermore, on the BCS side the effective chemical potential approaches the Fermi energy at low temperatures. Therefore, the fugacity is not small anymore, violating the validity of the virial expansion. Therefore, in the low-$T$ regime stretching from unitarity towards the BCS limit our measurements are particularly important and can serve as a benchmark for theoretical models.

The different regimes in the BCS-BEC crossover also	show up very clearly in Fig.~\ref{fig3}d), where we plot the contact as a function of temperature for $(k_F a_s)^{-1} = -0.5, \, 0$ and $1.5$.
On the BEC side at $(k_F a_s)^{-1} = 1.5$ the dimers dominate the contribution to the contact. For low enough temperatures, when  all atoms are bound in dimers, the contact is a constant (as a function of temperature). When $T \times k_B$ becomes comparable to the binding energy, the dimers become thermally unstable, break up and the contact starts decreasing (see Fig.~\ref{fig3}d) blue curves). On the BCS side and at unitarity where $(k_F a_s)^{-1} \leq 0$, no weakly-bound Feshbach molecular state exists. There, the decrease of the contact with increasing temperature is mainly due to overall decreasing atom density and to a breakdown of short-range pair correlations. Here, at low temperatures, our measurements for the contact strongly deviate from the results of the second-order quantum virial expansion (see Fig.~\ref{fig3}d) red and black curves).

\textit{Conclusion.---} In conclusion, we have precisely measured Tan's contact in the full phase diagram of the BCS-BEC crossover using photoexcitation of fermion pairs. Our results bridge the gap between the well-understood BCS and BEC regimes and are in line with recent measurements and calculations at unitarity. They extend previous measurements of Tan's contact to the finite temperature regime and are consistent with the quantum virial expansion for temperatures above $0.5 \, T_F$. 

For the future, we plan to extend our contact measurements to  homogeneous Fermi gases. It has been predicted (and measured at unitarity \cite{Mukherjee1}) \cite{Hu,Strinati1} that a sudden change in Tan's contact should appear at the critical temperature $T_C$ of superfluidity. Therefore, one could use such measurements to precisely map out $T_C$ within the BCS-BEC crossover. In the harmonically trapped system this sudden change is washed out due to the inhomogeneous density distribution in the trap. In addition, we also aim for studying the contact for systems of lower dimensionality or that feature spin imbalance. Here, probing pair correlations by measuring the contact might uncover the presence of the Fulde-Ferrell-Larkin-Ovchinnikov (FFLO) phase \cite{Hu2}.

~
\textit{Acknowledgement.---} We would like to acknowledge financial support by the German Research Foundation (DFG) through grant 382572300. We thank Hui Hu, Xia-Ji Liu, Jia Wang,  Sascha Hoinka, and Chris Vale  for valuable discussions. We thank Frederik Koschnick and Wladimir Schoch  for general assistance in the lab, and Markus Dei{\ss}, Dominik Dorer, Jinglun Li and Wolfgang Limmer for discussions and support.

\bibliographystyle{apsprl}

\renewcommand{\thefigure}{S\arabic{figure}}
\renewcommand{\theequation}{S\arabic{equation}}
\renewcommand{\tablename}{TABLE}
\renewcommand{\thetable}{S\arabic{table}}

\setcounter{equation}{0}
\setcounter{figure}{0}
\setcounter{table}{0}

\def\dd{\mathrm{d}}

\begin{widetext}

\begin{center}
   \Large{\textit{Supplemental Material for}}

\textbf{\large Precise photoexcitation measurement of Tan's contact in the entire BCS-BEC crossover} \\ \vspace{4mm}
\normalsize Manuel Jäger$^1$ and Johannes Hecker Denschlag$^1$\\
\textit{$^1$Institut für Quantenmaterie, Universität
 Ulm, 89069 Ulm, Germany}
 
\end{center}\vspace{-3mm}

\section{Photoexcitation scheme}
To induce an inelastic two-body loss in our strongly-interacting two-component gas of $^6$Li-atoms we use a laser to drive the optical transition from the most-weakly-bound bare molecular state,  $X^1 \Sigma_g^+, \vib = 38$, to the electronically excited, deeply-bound molecular state, $A^1 \Sigma_u^+, \vib' = 68$, as illustrated in figure \ref{figs1}. The two states have a large Franck-Condon overlap of 0.077 \cite{SHulet1}.
The bare molecular state $X^1 \Sigma_g^+, \vib = 38$ is collisionally admixed to the scattering state of atomic pairs via the broad Feshbach resonance at 832.2 G. 

\begin{figure}[h!]
	\includegraphics[width=0.43\textwidth] {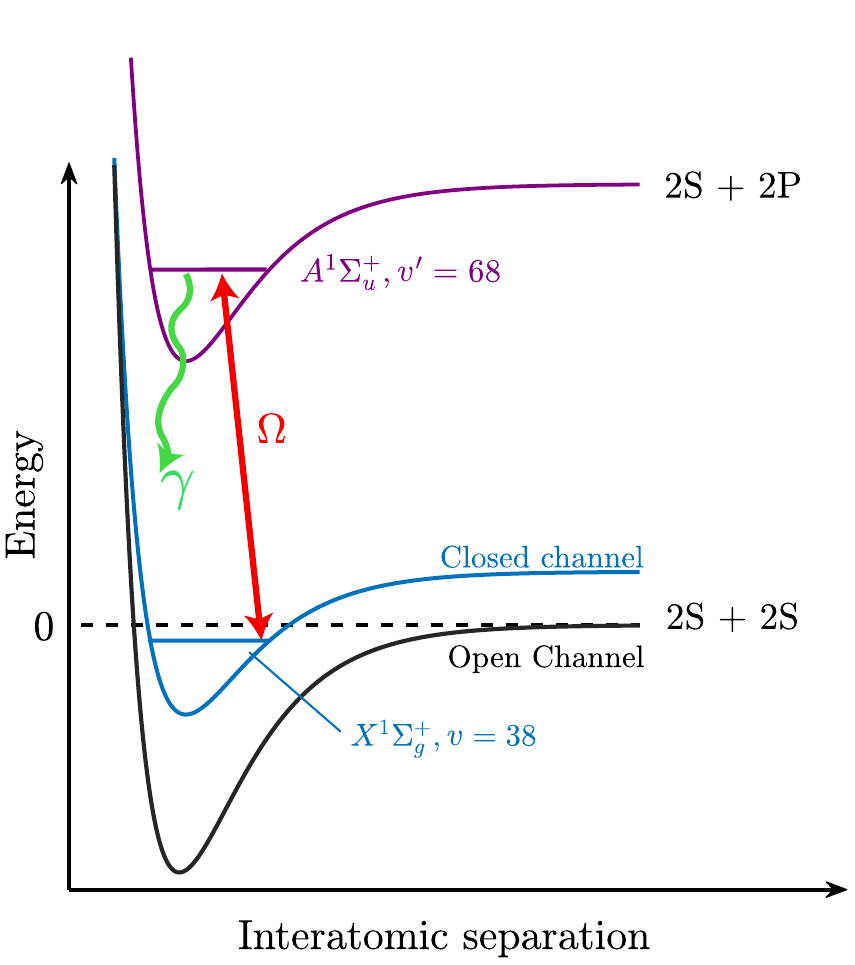}
	\caption{Photoexcitation scheme for the measurements of Tan's contact via two-body losses. We use a laser at 673 nm to couple the bare weakly-bound $X^1 \Sigma_g^+, \vib = 38$ state to the excited deeply-bound molecular state $A^1 \Sigma_u^+$  $\vib' = 68$  which has a lifetime $\gamma$. $\Omega$ is the Rabi frequency associated with this optical transition.}
	\label{figs1}
\end{figure}

\section{Tan's contact from the ground state energy expansions}
The total contact $\pazocal{I}$ for the harmonically trapped system can be derived from the (homogeneous) ground state energy expansions (see \cite{SWerner1}). Here, we briefly explain this approach.
According to Shina Tan \cite{STan2} the contact is related to the derivative of the systems energy $E$ with respect to the s-wave scattering length $a_s$ via
\begin{equation}
\frac{\dd E}{\dd (1/a_s)} =  - \frac{\hbar^2}{4 \pi m} \pazocal{I}.
\label{eq:tanoriginal}
\end{equation} 
$E$ is the total energy of the system including  kinetic energy, interaction energy, and the potential energy from the external trapping potential. This relation 
holds for a two-component Fermi gas at any temperature. 

At zero temperature  $T$  the total energy $E$ is known analytically at various regions of the BCS-BEC crossover. 
For a homogeneous Fermi gas in the BCS limit the energy per volume $V$ at zero temperature is given by the expansion 
\begin{align}
E/V = \frac{3}{5} \frac{\hbar^2 (k_F^\mathrm{hom})^2}{2m} n \left(1 + \frac{10}{9 \pi} k_F^\mathrm{hom} a_s +  0.1867 (k_F^\mathrm{hom} a_s)^2 + \dots \right),
\end{align}
where $n$ is the atom density and $k_F^\mathrm{hom} = (3 \pi^2 n)^{1/3}$ is the Fermi momentum of the homogeneous gas \cite{SWerner1, SBaker1, SLHY1}. The first three terms are the energy of the non-interacting Fermi gas, the Hartree-Fock mean field correction, and the (fermionic) Lee-Huang-Yang correction, respectively.  

Using relation \eqref{eq:tanoriginal} we obtain the contact density 
\begin{align}
C = \pazocal{I}/ V = 4 \pi^2 n^2 a_s^2 \left(1 + 1.049 k_F^\mathrm{hom} a_s + 0.2584 (k_F^\mathrm{hom} a_s)^2 + \dots \right).
\end{align}
At unitarity, the ground state energy density is 
\begin{align}
E/V = \frac{3}{5}\frac{\hbar^2 (k_F^\mathrm{hom})^2}{2 m} n \left(\xi + \frac{\zeta}{k_F^\mathrm{hom} a_s} + \dots \right),
\end{align}
where $\xi \approx 0.367$ \cite{SLi1} is the Bertsch parameter and $\zeta \approx 0.8$ \cite{SMukherjee1}. Therefore, the contact density is 
\begin{align}
C = \frac{6 \pi k_F \zeta}{5} n + \dots.
\end{align}
In the BEC limit of tightly bound dimers the energy density at zero temperature can be approximated by
\begin{align}
E/V = -n_d \frac{\hbar^2}{m a_s^2} + \frac{ 2\pi \hbar^2 a_{dd}}{2 m} n_d^2 \left(1  + 4.81 \sqrt{n_d a_{dd}^3} + \dots \right),
\label{eq:BEClim}
\end{align}
where the first term arises from the molecular binding energy for $a_s \rightarrow \infty$ and the next two terms from the mean field and the (bosonic) Lee-Huang-Yang correction, respectively \cite{SWerner1}. 
Here, $n_d = n/2$ is the dimer density and $a_{dd} = 0.6 a_s$ is the dimer-dimer scattering length \cite{SPetr1}. The contact density becomes 
\begin{align}
C = \frac{4 \pi n}{a_s} + 0.6 \pi^2 n^2 a_s^2 \left(1 + 12.03  \sqrt{n_d a_{dd}^3} + \dots \right).
\end{align}

Note that for finite scattering lengths $a_s$ the expression for the binding energy per molecule,
$$
 \frac{\hbar^2}{m a_s^2} ,
$$
in Eq. (\ref{eq:BEClim}) has to be modified by higher order corrections \cite{SPJ1, SGao1}. At a coupling of $(k_F^\mathrm{hom} a_s)^{-1} = 1 \, (2)$ these result in corrections of  $3.3\%$ ($6.8\%$) towards a larger contact density.

In a harmonic trapping potential, the atom density is position dependent and thus the contact changes locally. In the spirit of the local density approximation (LDA), the total contact of the system $\pazocal{I}$ is then obtained by integrating $C(\vec{r})$  over the trap volume $\pazocal{I} = \int \dd^3 r \,\, C(\vec{r})$. For this we follow the approach discussed in Appendix C of \cite{SWerner1}.\\

\section{Tan's contact from the EOS measurements}
In figure 2 of the main text we present a trap-integrated calculation of the contact $\pazocal{I}/N k_F$ based on the EOS measurements for a zero-temperature Fermi gas in the BCS-BEC crossover \cite{SNavon1}. In this section we briefly explain the procedure. 
In general, the total energy of a Fermi gas at zero temperature can be expressed as
\begin{equation}
E/V = \frac{3}{5} \frac{\hbar^2 (k_F^\mathrm{hom})^2}{2m} n \, \xi(\delta) - \Theta(a_s) n \frac{\hbar^2}{m a_s^2}, \label{eos2}
\end{equation}
where $\xi(\delta)$ can be viewed as a generalized Bertsch parameter relating the systems total energy to the Fermi energy at a given  interaction parameter $\delta$. $\Theta$ is the Heaviside step function and accounts for the existence of the Feshbach bound state for $a_s > 0$.
In the Supplementary Material of \cite{SNavon1} the function
\begin{equation}
    \xi(\delta) = \frac{h(\delta)-\frac{\delta}{3}h'(\delta)}{(h(\delta) - \frac{\delta}{5}h'(\delta))^{5/3}} \label{eos1}
\end{equation}
is defined. Here, $\delta$ is related to the coupling parameter $(k_F^\mathrm{hom} a_s)^{-1}$  through an implicit equation,
\begin{equation}
    (k_F^\mathrm{hom} a_s)^{-1} (\delta) = \frac{\delta}{(h(\delta) - \frac{\delta}{5}h'(\delta))^{1/3}} , \label{eos3}
\end{equation}
which can be solved numerically.\\
For the function $h(\delta)$ Padé-approximations were given for both the BEC ($a_s > 0$) and BCS ($a_s < 0)$ regime. The approximations, whose parameters were deduced from the EOS measurements, were made such that $h(\delta)$ is continuous at unitarity $(k_F^\mathrm{hom} a_s)^{-1} = 0$. The function $h'(\delta)$ denotes the derivative of $h(\delta)$.
By combining equations \eqref{eos2}, \eqref{eos1} and \eqref{eos3} we find the systems energy $E$ as a function of $(k_F^\mathrm{hom} a_s)^{-1}$. From this, the contact density $C$ across the BCS-BEC crossover can be determined as in the previous section. We then integrate over the trap volume as described in Appendix C of \cite{SWerner1} to get the total contact $\pazocal{I}$ for the harmonically trapped Fermi gas. Note that although the Padé-approximations for $h(\delta)$ are continuous at unitarity, they are not continuously differentiable there. As a result the calculated contact exhibits a kink there (see Fig. 2 of the main text).

\section{Photoexcitation-induced two-body decay laws in a harmonic trap}
According to equation (2) in the main text 
\begin{align}
-\frac{\dd N}{\dd t} = \frac{\hbar \, \pazocal{I}}{2 \pi m a_{bg} W}   \frac{\Omega^2 / (2 \gamma)}{\left[1 - a_{bg}/a_s\right]^{-2} + \left[\Omega^2 / (2 \gamma \, W)\right]^2 }, \label{eq:loss}
\end{align}
the expected atom loss via photoexcitation of pairs is proportional to the total contact $\pazocal{I}$ \cite{SHu}. If  the dependency of the contact on the atom number $N$ is known,  one can deduce the corresponding decay law for $N(t)$.

As calculated in \cite{SWerner1} in the zero temperature BCS limit the integration of the homogeneous contact density over the trap volume yields the proportionality $\pazocal{I} \propto k_F^3 N$ and therefore $\dot{N}/N \propto N^{1/2}$. At unitarity one finds $\pazocal{I} \propto k_F N$ and hence $\dot{N}/N \propto N^{1/6}$. In the BEC limit $\pazocal{I} \propto N$ such that $\dot{N}/N = \textrm{const}$. Here $k_F = \sqrt{2 m E_F}/\hbar$ is the Fermi momentum of the trapped gas, $E_F = k_B T_F = \hbar \bar{\omega} (3 N)^{1/3}$ is the Fermi energy, and $\bar{\omega}$ is the geometric mean of the trapping frequencies. 
Thus, these differential equations have the form
 $\dot{N}/N \propto N^{1/b}$,
 where $b = 2$ in the BCS limit, $b = 6$ at unitarity and $b \rightarrow \infty$ in the BEC limit.
 Their solutions, i.e. the decay laws, are given by 
 the power law
\begin{equation}
N(t) = N_0 / (1 + \Gamma_0 t / b)^b,
\label{eq:decay}
\end{equation}
with the initial decay rate $\Gamma_0$.

\section{Experimental parameters to the measurements shown in Fig. 1}
In Table \ref{tab:data}  we list the experimental parameters for the three curves shown in Fig. 1 of the main text.
\begin{table}[h]
    \centering
    \begin{tabular}{c|c|c|c|c|c|c|c}    
       \multirow{2}{*}{Magnetic field} & \multirow{2}{*}{Atom number} & Coupling & \multirow{2}{*}{Temperature} & Photoexcit. & Theo. & Fit  & Fit param. $b$ \\
        & & $(k_F a_s)^{-1}$ & &  laser power & param. $b$ & param. $b$ & $t <$ 250 ms\\
       \hline
       753 G & $5.9 \times 10^5$ & 1.5 & 0.07 $T_F$ & 6 $\upmu$W &$\infty$ & $> 20$ & $> 20$ \\
       832 G & $6.05 \times 10^5$  & 0 & 0.05  $T_F$ & 38 $\upmu$W & 6 & $3.9 \pm 0.9$ & $5.7 \pm 3.1$\\
       1078 G & $4.8 \times 10^5$ & -1.65 & 0.04  $T_F$ & 610 $\upmu$W & 2 & $1.6 \pm 0.2$ & $2.1 \pm 0.5$\\       
    \end{tabular}
    \caption{Experimental parameters for the measurement data shown in Fig. 1 of the main text. The values for atom numbers, couplings and temperatures are the initial values at $t = 0$.}
    \label{tab:data}
\end{table}
As mentioned there, we observe a deviation in the fit parameters $b$ (see Eq. (3)) compared to the theoretical predictions. We attribute this deviation to an increase of the atom cloud temperature during the photoexcitation pulse, especially for longer times ($>$ 400 ms). During the first 250 ms, however, the temperature stays rather constant within 5-10\%. If we only include these data points, the fitted parameters tend towards the theoretical values. However, the uncertainty  increases since we lower the number of data points for the fit. Note that after 400-500 ms, thermometry based on the atom density distributions becomes increasingly difficult as the density and therefore the signal decreases.

\section{Atom number, trapping frequency and magnetic field calibration}

For determining the atom number, we carefully calibrated our absorption imaging scheme at high magnetic fields. For this purpose we use a short laser pulse of 10 $\mu$s at an intensity of $I/I_S = 0.05$ to drive the quasi-closed transition to the $|2^2P_{3/2};m_J = -3/2>$ state, where $I_S = 2.54 \, \textrm{mW}/\textrm{cm}^2$ is the saturation intensity. During the imaging the atoms scatter photons. This accelerates them and leads to an increasing Doppler shift during imaging (see \cite{SHueck1}). In order to take this effect into account, we calibrated our imaging routine using a simple classical mechanical model. This allows for extracting the accurate atom number at any given intensity and pulse duration. With this, we can determine atom numbers with a typical uncertainty of 5$\%$, corresponding to an uncertainty in $k_F \propto N^{1/6}$ of 1$\%$. We further tested this calibration using calculations of the atom cloud density based on the equation of state (EOS) at unitarity \cite{SNasc1} and with a mean field model in the BEC regime as described in \cite{SDH1}. For this, we calculate the 2D column density and 1D line density for given atom numbers, trapping frequencies and temperatures and compare the calculated densities to the measured ones.

For determining the trapping frequencies, we perform either parametric heating by modulating the potential of the optical dipole trap or we observe the center of mass motion of the atom cloud after an initial small displacement of a few micrometers. Both methods give consistent results with approximately 2$\%$ uncertainty for the parametric heating method and 3$\%$ uncertainty for the center of mass motion measurement.
        
For measuring the magnetic fields precisely, we perform radio-frequency (RF) spectroscopy between the two lowest atomic hyperfine states. For this, we first use a short laser pulse to depopulate the $|F = 1/2, m_F = 1/2>$ state. We then apply a $50$ ms RF pulse and scan stepwise the RF frequency of $\approx 76.2$ MHz to find resonant population transfer to this state from the still populated  $|F = 1/2, m_F = -1/2>$ state. Using the Breit-Rabi formula we are able to determine the magnetic field with an uncertainty of about 0.5 G, corresponding to an uncertainty of $(k_F a_s)^{-1}$ of $<1 \%$. To assign a scattering length to the measured magnetic fields we use reference \cite{SHutson1}.

\section{Tan's contact within the quantum virial expansion}
To calculate the finite temperature contact we use Tan's adiabatic sweep theorem for the grand canonical ensemble 
\begin{equation}
\left(\frac{\partial \Omega_G}{\partial (1 / a_s)} \right) _{T,V,\mu} = - \frac{\hbar^2 C V}{4 \pi m}  \label{eq:sweep}
\end{equation}
and the quantum virial expansion (see \cite{SHu1}).
In the virial expansion, the grand canonical potential
\begin{equation}
\Omega_G = - 2k_B T V / \lambda_{dB}^3 \left(z + b_2 z^2 + b_3 z^3 + \dots\right) \label{eq:virial}
\end{equation}
is expanded as a series expansion in the fugacity $z =\exp(\mu / k_B T )$ where $b_n$ are the virial coefficients, $V$ is the volume and $\lambda_{dB}$ is the thermal de Broglie wavelength \cite{SXLiu}. The virial coefficients are functions of $\lambda_{dB}$ and the scattering length $a_s$. Using equations \eqref{eq:sweep} and \eqref{eq:virial} we find the homogeneous contact density
\begin{equation}
 C_{QV} = \frac{16}{\pi^2 \lambda_{dB}^4} \left(c_2 z^2 + c_3 z^3 + \dots \right)
\end{equation}
for a given temperature $T$, chemical potential $\mu$ and scattering length where $c_n = \partial b_n / \partial (\lambda_{dB} / a_s)$.
Using the local density approximation, one can calculate the total contact for the harmonically trapped Fermi gas with trap frequency $\omega$ by replacing the chemical potential $\mu$ with a local chemical potential $\mu(r) = \mu_0 - \frac{1}{2} m \omega^2 r^2$ and then integrating

\begin{figure}[h!]
	\includegraphics[width=0.50\textwidth] {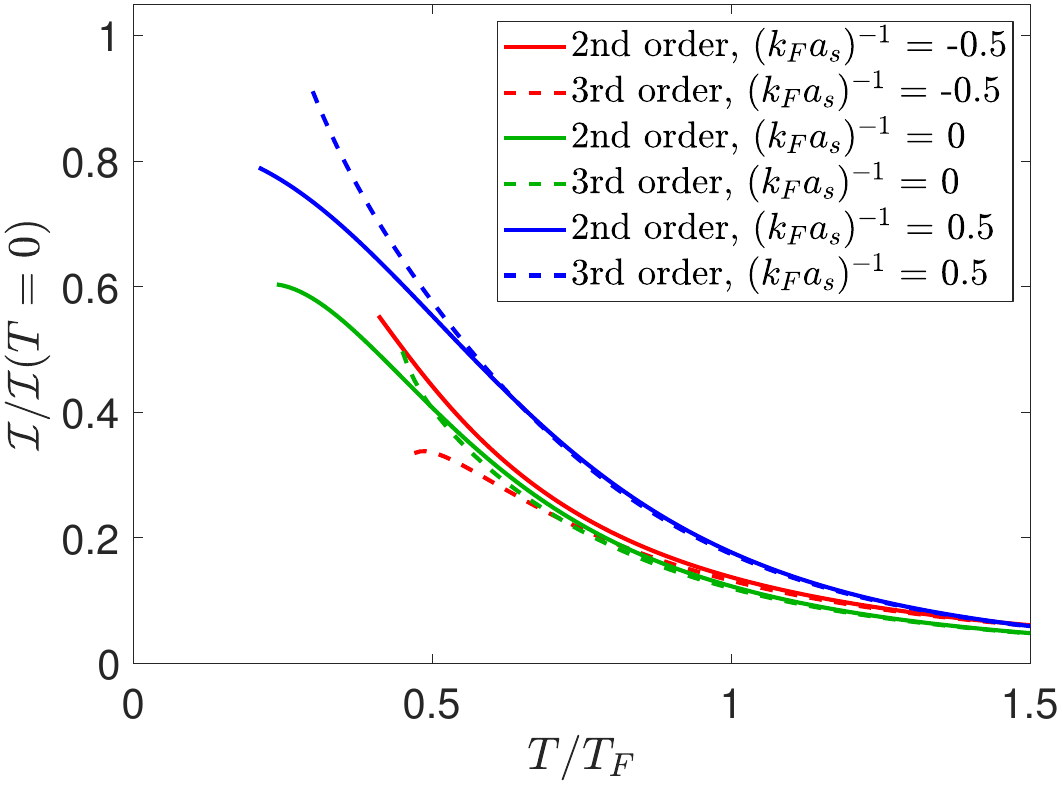}
	\caption{Contact $\pazocal{I}$ of a harmonically trapped Fermi gas, as calculated with the second and third order virial expansions for couplings $(k_F a_s)^{-1} = (-0.5, 0, 0.5)$. We normalized each contact with a respective zero-temperature value, taken from our interpolation of the zero-temperature measurement as shown in figure 2 of the main text.}
	\label{figs2}
\end{figure}

\begin{align}
\pazocal{I}_{QV} = \int \dd^3r \,\, C(r) &= \frac{16}{\pi^2 \lambda_{dB}^4} \int \dd^3r \,\, \left( c_2 z^2(r) + c_3 z^3(r) + \dots \right)  \\
 &= \frac{16}{\pi^2 \lambda_{dB}} \left( \frac{k_B T} {\hbar \omega} \right)^3 \left[ \frac{c_2}{2^{3/2}} z_0^2  +  \frac{c_3}{3^{3/2}} z_0^3 + \dots \right]. \label{eq:trapc}
\end{align}
Here, $\mu_0$ is the chemical potential in the center of the trap and $z_0 = \exp(\mu_0 / k_B T)$ is the corresponding fugacity.
In a harmonically trapped Fermi gas the chemical potential in the trap center is usually not known. What is known however is the total atom number $N$. Starting from the thermodynamic relation $n = - \frac{1}{V} \frac{\partial \Omega_G}{\partial \mu}$ and using again the local density approximation one finds
\begin{align}
N = \int \dd^3r \,\, n(r) &= 2 \left(\frac{k_B T}{\hbar \omega}\right)^3 \left[z_0  + \frac{b_2}{\sqrt{2}} z_0^2 + \frac{b_3}{\sqrt{3}} z_0^3 + \dots \right] \\
 &= 2 \left(\frac{k_B T}{\hbar \omega}\right)^3 \left[z_0  + \frac{b_2^{(0)} + \Delta b_2}{\sqrt{2}} z_0^2 + \frac{b_3^{(0)} + \Delta b_3}{\sqrt{3}} z_0^3 + \dots \right]\\
 &= 2 \left(\frac{k_B T}{\hbar \omega}\right)^3 \left[z_0  - \frac{z_0^2}{2^3} +  \frac{z_0^3}{3^3} + \dots + \frac{\Delta b_2}{\sqrt{2}} z_0^2 + \frac{\Delta b_3}{\sqrt{3}} z_0^3 + \dots  \right]\\
 &= 2 \left(\frac{k_B T}{\hbar \omega}\right)^3 \left[-\mathrm{Li}_3 (-z_0) \dots + \frac{\Delta b_2}{\sqrt{2}} z_0^2 + \frac{\Delta b_3}{\sqrt{3}} z_0^3 + \dots  \right], \label{eq:chemN}
\end{align}
where we separated the virial coefficients in parts $\Delta b_n$ that take into account the n-body interactions (e.g. scattering properties, bound states,...) and parts that account for  quantum statistics $b_n^{(0)} = \frac{(-1)^{n+1}}{n^{5/2}}$. The infinite sum \mbox{$z_0  - z_0^2/2^3 +  z_0^3/3^3 + \dots$} can be identified as the polylogarithm function $\mathrm{Li}_3(-z_0)$ of degree 3  and argument $-z_0$. The polylogarithm function is related to the integral of the Fermi-Dirac distribution function.\\

Equation \eqref{eq:chemN} can be numerically solved to determine $\mu_0$ for a given atom number $N$, temperature $T$, scattering length $a_s$ and trap frequency $\omega$. With the chemical potential one can finally calculate the contact of the trapped Fermi gas using equation \eqref{eq:trapc}.\\
The second order virial coefficient is analytically known and reads \cite{SXLiu,SBeth} 
\begin{align}
b_2 = b_2^{(0)} + \Delta b_2 = \frac{-1}{2^{5/2}} + \sqrt{2} \left[\Theta(a_s) \mathrm{e}^{\lambda_{dB}^2/ 2 \pi a_s^2} - \frac{1}{2} \mathrm{sgn}(a_s) \left(1 - \mathrm{erf} \left[ \sqrt{\lambda_{dB}^2/ 2 \pi a_s^2} \right]   \right) \mathrm{e}^{\lambda_{dB}^2/ 2 \pi a_s^2} \right],
\end{align}
where $\Theta(\dots)$ is the Heavyside step function, $\mathrm{sgn(\dots)}$ is the sign function and $\mathrm{erf}(\dots)$ is the error function. 
We also performed calculations of the contact with the third order virial expansion. For this, we extracted $b_3$ from \cite{SLeyr}. Above $(k_F a_s)^{-1} = 1.5$ the second and third order calculation give the same results for the contact. As also observed in \cite{SHu1}, we find that the second and third order results at unitarity start to deviate from each other at temperatures lower than $\approx 0.5\, T_F$. Additional calculations for couplings $(k_F a_s)^{-1} = (-0.5, 0.5)$ are shown in figure \ref{figs2}. When we go deeper into the BCS regime the second and third order results start to deviate at temperatures even higher than 0.5 $T_F$. This is because in the BCS regime the fugacity $z_0 = \exp(\mu_0 / k_B T)$ is not small anymore as the chemical potential $\mu_0$ is positive and approaches the Fermi energy for $(k_F a_s) \rightarrow -\infty$. As a result the virial expansion loses its validity and the higher order expansions do not converge.\\

\section{Closed-Channel fraction}

In the vicinity of the Feshbach resonance the scattering wavefunction $\Psi$ of an atom pair has a closed-channel admixture of the  bare highest bound molecular state $X^1 \Sigma_g^+ (\vib = 38)$.
\begin{figure}[h!]
 \includegraphics[width=0.80\textwidth] {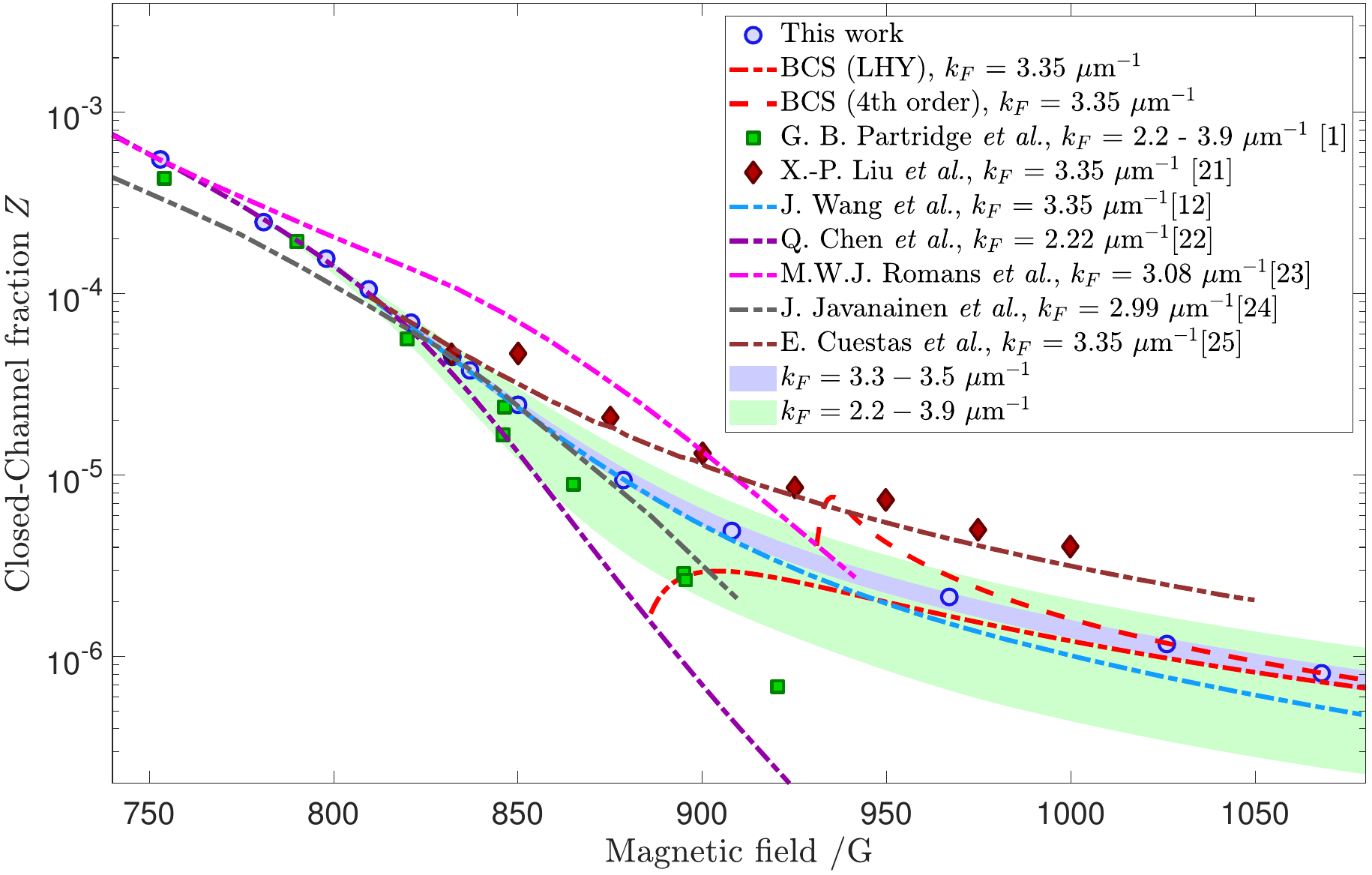}
 \caption{Closed-channel fraction $Z$ as a function of magnetic field, taken from various studies. Blue circles are extracted from our measured contact data using equation \eqref{eq:relationcontactz}. The blue (green) shaded area is a calculation of the closed-channel fraction based on these results for the range of $k_F$ from $3.2$ to $3.5\,\upmu$m$^{-1}$ ($2.2$ to $3.9\,\upmu$m$^{-1}$). Data from other studies (calculations and measurements) have plot symbols as indicated by the legend. This includes photoexcitation measurements from X.-P. Liu  \textit{et al.} \cite{SJWP} and from G.B. Partridge \textit{et al.} \cite{SHulet1}, and calculations by J. Wang \textit{et al.} \cite{SHu}, Q. Chen \textit{et al.} \cite{SLevin1}, M.W.J. Romans \textit{et al.} \cite{SRomans}, J. Javanainen \textit{et al.} \cite{SJavanainen}, and E. Cuestas \textit{et al.} \cite{SCuestas}. We further show calculations based on the BCS ground-state energy expansions as in figure 2 of the main text.}
 \label{figs3}
\end{figure}
It can be written as $\Psi = \sqrt{Z} \, \Psi_\textrm{closed} + \sqrt{1 - Z} \, \Psi_\textrm{open}$ where $Z$ is the so-called closed-channel fraction. For a weak probe laser intensity ($\Omega^2 \ll \gamma W$) this quantity is directly linked to Tan's two-body contact via \cite{SWerner1}
\begin{equation}
Z = \frac{\pazocal{I}}{N k_F}  \frac{\hbar k_F}{2 \pi m a_{bg} W} \left[1 - \frac{a_{bg}}{a_s} \right]^2,
\label{eq:relationcontactz}
\end{equation}
and has been experimentally and theoretically investigated by several groups \cite{SHulet1,SJWP,SHu,SLevin1,SRomans,SJavanainen,SCuestas}. Here $W$ is the width of the Feshbach resonance. It is therefore natural to compare all these data, see Fig.~\ref{figs3}. However, since the closed-channel fraction is not a normalized quantity it can only be compared directly for measurements and calculations with similar parameters $W, k_F, m, a_{bg}, a_s$. Apart from some variations in the Fermi momentum $k_F$ the parameters are indeed the same for the different data sets. Therefore, a  quantitative comparison is approximately possible.\\
To calculate the magnetic field dependence of the closed channel fraction for any given Fermi momentum $k_F$  we use the interpolation of our measurements of the total contact as presented in figure 2 of the main text. The measured quantity $\pazocal{I}/N k_F$ only depends on $(k_F a_s)^{-1}$ to a good approximation. Thus, for a given Fermi momentum $k_F$ and magnetic field $B$ to which we assign a corresponding scattering length $a_s$, we can determine $\pazocal{I}/N k_F$ from our interpolation. Using this result and equation \eqref{eq:relationcontactz}, we can then calculate the closed-channel fraction $Z$. We use this procedure to calculate the blue and green shaded areas in figure \ref{figs3}.\\
With the blue shaded area we illustrate a lower and upper bound for the closed channel fraction deduced from our zero-temperature contact measurements where the atom number ranges from $4.8 \times 10^5 - 6.5 \times 10^5$ (corresponding to $k_F = 3.3 - 3.5\,\upmu$m$^{-1}$). We further plot the closed channel fraction calculated for $k_F = 2.2 - 3.9\,\upmu$m$^{-1}$. This allows us to compare our measurements to the measurements presented in \cite{SHulet1} where the Fermi momentum $k_F$ ranges from $2.2$ to $3.9\,\upmu$m$^{-1}$ and to the calculation presented in \cite{SLevin1} for $k_F = 2.2\,\upmu$m$^{-1}$.
The plot shows that up to this point there is still a large discrepancy between different measurements and theories on the BCS side of the Feshbach resonance. This highlights the importance of high precision measurements in this regime.

\section{Quasi 1D Projection}
We make an interesting observation, when plotting our measurements from Fig. 3 of the main text in a three-dimensional coordinate system, as shown in Fig. \ref{figs4} a). 
By rotating the graph by the azimuthal and polar angles, we can align the data points so that, at a particular line of sight, they approximately fall onto a single universal 1D curve, as shown in Fig \ref{figs4} b).  
We parametrize the line of sight by two angles ($\phi,\theta)$. The azimuthal angle $\phi$ is the angle between the line of sight and the negative y-axis (temperature) and the angle of elevation $\theta$ is the angle between the line of sight and the x-y (coupling-temperature) plane. We find that for angles of $\phi = -24^\circ$ and $\theta = -8^\circ$ the convergence of data onto a single line is best. This behavior, however,  only works in a limited realm of $T/T_F \lesssim 2$ and $-1 \lesssim 1/k_F a_s \lesssim 2$.
Second order quantum virial calculations at high temperatures and calculations using the ground state energy expansions of the BCS and BEC limits show this.

\begin{figure}[ht]  
\includegraphics[width=.95\linewidth, trim={2.2cm 19.9cm 2cm 1.9cm},clip]{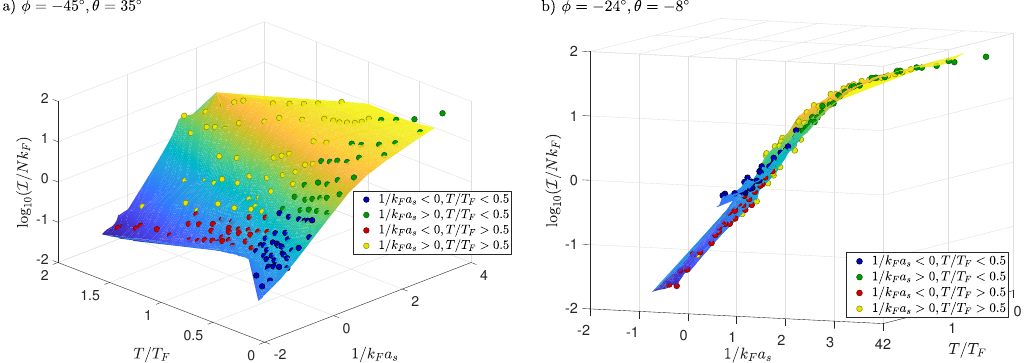}   
\caption{a) Measured contact values (dots) in a 3-dimensional representation. The colored plane is a linear interpolation of the data as shown in Fig. 3 a) of the main text. By changing the  viewer's line of sight onto the dataset, the data approximately collapse onto a single curve, see b). The angles $\phi$ and $\theta$ specify the line of sight (see text). In order to better distinguish the measured data points belonging to  the BEC and BCS regimes and to low and high temperatures, we used different colors for the plot symbols (see legend).} 
 \label{figs4}   
\end{figure}


\bibliographystyle{apsprl}

\end{widetext}
\end{document}